\documentclass[twocolumn,superscriptaddress,secnumarabic,
amssymb,amsmath,nobibnotes,aps,prd,showkeys,showpacs,nofootinbib]{revtex4}%
\usepackage{graphicx}
\usepackage{epsf}
\usepackage{bm}
\usepackage{amsmath}
\usepackage{amsfonts}
\usepackage{amssymb}
\usepackage{epstopdf}
\usepackage{natbib}
\usepackage{color}
\usepackage{float}
\providecommand{\U}[1]{\protect\rule{.1in}{.1in}}

\newcommand{\be}{\begin{equation}}
\newcommand{\ee}{\end{equation}}

\newcommand{\mincir}{\raise
-3.truept\hbox{\rlap{\hbox{$\sim$}}\raise4.truept\hbox{$<$}\ }}
\newcommand{\magcir}{\raise
-3.truept\hbox{\rlap{\hbox{$\sim$}}\raise4.truept\hbox{$>$}\ }}

\newtheorem{remark}{Remark}[section]

\begin{document}

\title{Gravitational production of  superheavy baryonic and dark matter in quintessential inflation: nonconformally coupled case}

\author{Jaume   Haro}
\email{jaime.haro@upc.edu}
\affiliation{Departament de Matem\`atiques, Universitat Polit\`ecnica de Catalunya, Diagonal 647, 08028 Barcelona, Spain}


\author{Llibert Arest\'e Sal\'o}
\email{llibert.areste-salo@tum.de} 
\affiliation{Departament de Matem\`atiques, Universitat Polit\`ecnica de Catalunya, Diagonal 647, 08028 Barcelona, Spain}
\affiliation{TUM Physik-Department, Technische Universit{\"a}t M{\"u}nchen, James-Franck-Str.1, 85748 Garching, Germany}

\thispagestyle{empty}

\begin{abstract}
The gravitational production of superheavy dark matter  is studied in the context of quintessential inflation. The superheavy particles, whose decay products  are baryonic matter and are responsible for the reheating of  the universe after the end of the inflationary period, are not conformally coupled with gravity. On the contrary,  dark matter is assumed to be conformally coupled with gravity. We show that the viability of these scenarios requires the mass of the superheavy dark matter to be 
greater than  $8\times 10^{15}$ GeV.
\end{abstract}

\vspace{0.5cm}

\maketitle
\section{Introduction}
Quintessential inflation, which was addressed for the first time by
 Peebles and Vilenkin (PV) in \cite{pv}, is an attempt to unify  inflation and quintessence via a single scalar field whose potential
allows inflation at early times while at late time provides quintessence (see for instance \cite{dimopoulos} and references therein).
A remarkable property of the  PV model  is that it contains an abrupt phase transition from inflation to  kination 
(a regime where all the energy density of the inflation turns into kinetic), where   the adiabatic regime is broken and, thus, particles could be gravitationally created \cite{ford, Damour}. This leads to the possibility to explain the abundance of dark matter through the gravitational production of superheavy particles 
during the phase transition in quintessential inflation \cite{hashiba, hashiba1}, or  during the oscillations of the inflaton field 
in standard inflation \cite{kolb1, ema,kolb2}.

The potential of the model presented here depends on two parameters which are determined using observational data: one with the observational value of the power spectrum of scalar perturbations and the other one taking into account that the ratio of the energy density of the scalar field to the critical energy density at the present time is approximately $0.7$. Moreover, this potential is obtained matching a Starobinsky inflationary-type potential with the inverse power law potential used in \cite{pv}. The former one leads to theoretical values of the spectral index and the ratio of tensor to scalar perturbations agreeing with the recent observational data provided by the Planck's team \cite{planck18}, and the second one is responsible for the current cosmic acceleration.

Since the potential has an abrupt phase transition at the end of the inflationary phase,  we will consider  the gravitational production of  two kinds of superheavy particles: $X$-particles, nonconformally coupled with gravity,
whose energy density after their decay into baryonic light particles and later thermalization of decay products will dominate the energy density of the scalar field in order to match with the Hot Big Bang (HBB), and dark $Y$-particles, conformally coupled with gravity,  which  are only gravitationally interacting massive particles (GIMP).
We will show that  the quintessential inflation model  presented in this work preserves the Big Bang Nucleosynthesis (BBN) success, in the sense that the overproduction of Gravitational Waves (GWs) does not disturb the BBN   for $X$-particles and $Y$-particles with masses in the range of { $10^{15}-10^{17}$  GeV  and $10^{16}-10^{18}$ GeV} respectively,  leading to a maximum reheating temperature in the TeV regime.

The paper is organized as follows: In Section II we present our quintessential inflation model based on a Starobinsky Inflation-type potential matched with a quartic inverse
power law potential. Section III is devoted to the calculation of the energy density of the superheavy produced particles and to give viable bounds for the reheating temperature
and for the masses of $X$ and $Y$ particles. In Section IV numeric calculation has been performed in order to show the viability of the model at the present time and its future
evolution. Finally, we present the conclusion of the work  in Section V.

\

The units used throughout the paper are $\hbar=c=1$ and   the reduced Planck's mass  is denoted by 
$M_{pl}\equiv \frac{1}{\sqrt{8\pi G}}\cong 2.4\times 10^{18}$ GeV.

\section{The quintessential inflation model}
It is well-known that in quintessential inflation the number of e-folds from the pivot scale exiting the Hubble radius  to the end of inflation is greater than $60$. 
For this reason,
in order that  the theoretical values of the spectral index and the ratio of tensor to scalar perturbations enter  in their marginalized joint confidence contour
in the plane
$(n_s,r)$ at $2\sigma$ C.L. for the 
Planck2018 TT, TE, EE + low E+ lensing + BK14 + BAO likelihoods \cite{planck18},  
we have changed the quartic inflationary potential of the original  
 PV quintessential inflation model \cite{pv} by a Starobinsky-type potential in the Einstein Frame (EF)  \cite{riotto,aho} (also named {\it Higgs Inflation} \cite{martin}), obtaining:
\begin{eqnarray}\label{pv}
V(\varphi)=\left\{\begin{array}{ccc}
\lambda M_{pl}^4\left(1-e^{\sqrt{\frac{2}{3}}\frac{\varphi}{M_{pl}}}+\frac{M^2}{M_{pl}^2}\right)^2& \mbox{for}& \varphi\leq 0\\
\lambda\frac{M^8}{\varphi^4+M^4} &\mbox{for}& \varphi\geq 0,
\end{array}\right.
\end{eqnarray}
where $\lambda$ is a dimensionless parameter which we will calculate right now and, as we will show in Section IV, $M\cong 2.6 \times 10^5$ GeV is a small mass.

\begin{remark} To show the equivalence of $R^2$-gravity in the Jordan Frame (JF) and the first piece of the potential (\ref{pv}) in the EF, we consider, in the flat 
Friedmann-Lema{\^\i}tre-Robertson-Walker (FLRW) metric, the Lagrangian of $R^2$-gravity in the JF (see for instance \cite{odintsov})
\begin{eqnarray}
{\mathcal L}_{JF}=\frac{M_{pl}^2}{2}\left( R+\alpha R^2\right)a^3,
\end{eqnarray}
where $\alpha$ is a positive parameter with dimension of $M_{pl}^{-2}$.

To work in the EF, we perform the change of variable \cite{aho}
\begin{eqnarray}
\tilde{a}=a\sqrt{1+2\alpha R},\quad d\tilde{t}=dt\sqrt{1+2\alpha R}.
\end{eqnarray}

Then, the Lagrangian in the EF becomes
\begin{eqnarray}
{\mathcal L}_{EF}=\left(\frac{M_{pl}^2}{2} \tilde{R}+\frac{1}{2}(\tilde{\varphi}')^2-V(\tilde{\varphi})
\right)\tilde{a}^3,
\end{eqnarray}
where $'$ denotes the derivative with respect to $\tilde{t}$, the Ricci scalar in the EF is $\tilde{R}=6(\tilde{H}+2\tilde{H}^2)$, with 
$\tilde{H}=\tilde{a}'/\tilde{a}$, and the relation between both frames is given by
\begin{eqnarray} 
\tilde{\varphi}=-\sqrt{\frac{3}{2}}M_{pl}\ln(1+2\alpha R),\quad V(\tilde{\varphi})=\frac{\alpha R^2M_{pl}^2}{2(1+2\alpha R)^2}.
\end{eqnarray}

Therefore, since $1+2\alpha R= e^{-\sqrt{\frac{2}{3}}\frac{\tilde{\varphi}}{M_{pl}}}$, we conclude that  $R^2$-gravity in the JF
is equivalent to General Relativity (GR) in the EF when the potential is given by
\begin{eqnarray}
V(\tilde{\varphi})=\frac{M_{pl}^2}{8\alpha}\left( 1-e^{\sqrt{\frac{2}{3}}\frac{\tilde{\varphi}}{M_{pl}}}\right)^2.\end{eqnarray}

On the other hand, the tail of the potential $(\varphi>0)$, 
which is motivated by SUSY QCD \cite{SUSY}, 
is the same used by Peebles and Vilenkin in his seminal paper [1], and has been studied in many papers dealing with quintessence such as \cite{rp, albrecht}.

\end{remark}

\

In this model,  the kination phase starts at $\varphi_{kin}\cong 0$.  Thus,  to obtain the value of the Hubble parameter at that time, namely 
$H_{kin}$,
first of all we calculate the slow roll parameters:
Denoting by  $\epsilon_*=\frac{M_{pl}^2}{2}\left(\frac{V_{\varphi}(\varphi_*)}{V(\varphi_*)}  \right)^2$ and 
$\eta_* ={M_{pl}^2}\frac{V_{\varphi\varphi}(\varphi_*)}{V(\varphi_*)}$ the values of the slow roll parameters  and by $\varphi_*$ the value of the scalar field  when the pivot scale exits the Hubble radius,  since the mass $M$ satisfes $M\ll M_{pl}$, 
 one has $
\epsilon_*\cong \frac{4}{3}e^{2\sqrt{\frac{2}{3}}\frac{\varphi_*}{M_{pl}}}  $
$\eta_*=    -\frac{4}{3}e^{\sqrt{\frac{2}{3}}\frac{\varphi_*}{M_{pl}}} ,$
and thus, the spectral index is given by \cite{btw}
\begin{eqnarray}
1-n_s\cong 6\epsilon_*-2\eta_*\cong \frac{8}{3}e^{\sqrt{\frac{2}{3}}\frac{\varphi_*}{M_{pl}}},
\end{eqnarray}
meaning that
\begin{eqnarray}
 \varphi_*\cong \sqrt{\frac{3}{2}}M_{pl}\ln\left(\frac{3}{8} (1-n_s)\right).
\end{eqnarray}


On the other hand, the observational estimation of the power spectrum of the scalar perturbations when the pivot scale leaves the Hubble radius is ${\mathcal P}_{\zeta}\cong \frac{H_*^2}{8\pi^2M_{pl}^2\epsilon_*}\sim 2\times 10^{-9}$ \cite{btw}. 
Since during the slow roll regime the kinetic energy density is negligible compared with the potential one, we will have 
$H_*^2\cong \frac{\lambda}{3} M_{pl}^2$, and using the relation $\epsilon_*=\frac{3}{16}(1-n_s)^2$ one gets
\begin{eqnarray} 
\lambda\sim 9\pi^2(1-n_s)^2\times 10^{-9}.
\end{eqnarray}

Taking into account that the observational value of the spectral index is $n_s=0.968\pm 0.006$ \cite{Planck}, if one chooses  its central value one gets
\begin{eqnarray} \lambda=9\times 10^{-11} \quad  \mbox{and} \quad
\varphi_*\cong -5.42 M_{pl}. 
\end{eqnarray}

Then, once we have these quantities we can solve numerically the conservation equation
\begin{eqnarray}\label{KG}
\ddot{\varphi}+3
\sqrt{\frac{\frac{\dot{\varphi}^2}{2}+V(\varphi)}{3M_{pl}^2}}\dot{\varphi}+V_{\varphi}=0
\end{eqnarray}
with initial conditions $\varphi_*=-5.42 M_{pl}$ and $\dot{\varphi}_*=0$ (obviously, one can choose other similar initial conditions and the result has to be practically the same because the inflationary dynamics are that of an attractor). 

 Using event-driven integration with an ode RK78 integrator one gets
$\dot{\varphi}_{kin}=3.54\times 10^{-6} M_{pl}^2$, and thus
 \begin{eqnarray}
  H_{kin}=\frac{\dot{\varphi}_{kin}}{\sqrt{6} M_{pl}}\cong 1.44\times 10^{-6} M_{pl},
  \end{eqnarray}
and 
\begin{eqnarray}\label{8}
\rho_{\varphi, kin}\cong  6.26\times 10^{-12}M_{pl}^4.
\end{eqnarray}

To end this section, let's calculate the number of e-folds between the time when $\varphi=\varphi_*$ and $\varphi=\varphi_{END}$ (i.e. the end of inflation) provided by our model

\begin{eqnarray}
N=\int_ {t_*}^{t_{END}}Hdt=\frac{1}{M_{pl}}\int_{\varphi_*}^{\varphi_{END}}\frac{1}{\sqrt{2\epsilon}}d\varphi
\end{eqnarray}

So, using the value of $\varphi_*$ above, that $\epsilon\cong\frac{4}{3}\left(\frac{s}{1-s}\right)^2$, where $s=e^{\sqrt{\frac{2}{3}}\frac{\varphi}{M_{pl}}}$, and that $s_{END}\cong-3+2\sqrt{3}$ (which corresponds to $\epsilon_{END}=1$), one gets that
\begin{eqnarray}
N\cong \frac{3}{4}\left(\frac{8}{3(1-n_s)}+\frac{1}{3-2\sqrt{3}}+\ln\left(\frac{3}{8}\frac{n_s-1}{3-2\sqrt{3}}\right) \right),
\end{eqnarray}
which leads to $41.34\leq N\leq 95.29$ for the values of $0.956\leq n_s\leq 0.98$ within its $2\sigma$ C.L. In particular,
at $1\sigma$ C.L., i.e., 
 for the values $0.969\leq n_s\leq 0.975$,  the expected number of e-folds in quintessential inflation, is  between $60$ and $75$.



\section{Reheating via gravitational particle production}
Since the second derivative of the potential (\ref{pv}) is discontinuous at $\varphi=0$,  
from the conservation equation one can see that the third temporal derivative of the inflation field is discontinuous at the beginning of kination, and using the Raychaudhuri 
equation $\dot{H}=-\frac{\dot{\varphi}^2}{2M_{pl}^2}$ one can deduce that
at the beginning of kination the third derivative of the Hubble parameter is discontinuous,  enhancing the particle production as discussed in \cite{kolb}.
Then, 
in order that vacuum polarization effects do not disturb the dynamics of the $\varphi$-field, the mass of the
superheavy particles, produced gravitationally, must be greater than $10^{15}$ GeV,  where we have assumed that the beginning of inflation
occurs at GUT scales, that is, when the Hubble parameter is of the order of $10^{14}$ GeV (see for instance \cite{hyp}). 
Therefore,  for the $Y$-particles, which we assume to be conformally coupled with gravity,  since $m_Y\gg H$ one can safely use the WKB approximation (see section $2$ of \cite{bunch} for a detailed explanation) to calculate the $\beta$-Bogoliubov coefficient of the $k$-mode \cite{hap1}, leading for our model to
 \begin{eqnarray}
 |\beta_k(\tau)|^2\cong 
 \frac{m_Y^4a_{kin}^{12}(\dddot{H}(\tau_{kin}^-)-\dddot{H}(\tau_{kin}^+))^2}{1024\omega^{12}_k(\tau_{kin})}, 
\end{eqnarray}
where $\tau_{kin}$ denotes the beginning of the kination in conformal time,  $\omega_k(\tau)=\sqrt{k^2+a^2(\tau)m_Y^2}$ is the time dependent frequency of the $k$-mode and the third derivative of the Hubble parameter is evaluated on the right $(+)$  and  on the left  $(-)$ of $\tau_{kin}$.

 \

 \begin{remark}
 In \cite{hps} the calculation of the $\beta$-Bogoliubov coefficient was done using the well-known diagonalization method \cite{Grib,Zeldovich},  and the
 importance of the discontinuity of some derivative (in our case the second one) of the potential at the phase transition is pointed out. In fact, the greater the order of the discontinuous derivative is, the less the number density of superheavy gravitationally produced particles \cite{hap} is, which is in agreement with \cite{kolb}. So, for a smooth phase transition the production of superheavy particles would be suppressed and its energy density  would be abnormally small, meaning that in such a model the reheating is impossible via gravitational production of superheavy particles and, thus, other mechanisms of reheating, such as {\it "instant preheating"} \cite{fkl,fkl1}, must be invoked.
 \end{remark}

 \

On the contrary, for the $X$-particles, 
which are nonconformally coupled with gravity, we have that the $k$-mode satisfies the equation \cite{bunch}
\begin{eqnarray}
\chi''_k+\Omega_k^2\chi_k=0,
\end{eqnarray}
where $\Omega_k^2=\omega_k^2+(\xi-\frac{1}{6})a^2 R$, being
$\xi$ the coupling constant, $\omega_k(\tau)=\sqrt{k^2+a^2(\tau)m_X^2}$ and $R$ the Ricci scalar. 
At this point, one has to note that the WKB is a perturbative approximation which holds when
$m_X\gg |\xi-\frac{1}{6}|R$, and  thus, since at the GUT scales one has $R\sim 10^{29} \mbox{ GeV}^2$ so that the mass $m_X$ is far from the Planck's mass, one has to choose $|\xi-\frac{1}{6}|\leq 1$, and the square of the $\beta$-Bogoliubov is given by
 \begin{eqnarray}
 |\beta_k(\tau)|^2\cong \frac{9  (\xi-1/6)^2a_{kin}^8(\dddot{H}(\tau_{kin}^-)-\dddot{H}(\tau_{kin}^+)^2}{32\omega_k^8(\tau_{kin})}.
\end{eqnarray}



 Therefore, taking into account that
 \begin{eqnarray}
 (\dddot{H}(\tau_{kin}^-)-\dddot{H}(\tau_{kin}^+))^2=\dot{\varphi}_{kin}^4\left(\frac{V_{\varphi\varphi}(0^-)}{M_{pl}^2}\right)^2
 \nonumber\\
 =\frac{16\lambda^2}{9}\dot{\varphi}_{kin}^4
  \end{eqnarray}
  and the fact that the energy density of $A$-particles, with  $A=X, Y$, is given by
  {\begin{eqnarray}\label{particleproduction}
\rho_A(\tau)\cong \frac{m_A}{2\pi^2a^3(\tau)}\int_0^{\infty} k^2 |\beta_k(\tau)|^2 dk
\end{eqnarray}  
  before the decay of the $X$-particles, its energy density evolves as
{\begin{eqnarray}\label{particleproductionX}
\rho_X(\tau)
\cong \frac{\lambda^2}{128\pi}\left( \xi-\frac{1}{6} \right)^2\left( \frac{\dot{\varphi}_{kin}}{m_X}\right)^4\left( \frac{a_{kin}}{a(\tau)} \right)^3,\end{eqnarray}}
and the one of the $Y$-particles evolves as
{\begin{eqnarray}\label{particleproductionX}
\rho_Y(\tau)
\cong \frac{7\lambda^2}{589824\pi}\left( \frac{\dot{\varphi}_{kin}}{m_Y}\right)^4\left( \frac{a_{kin}}{a(\tau)} \right)^3.
\end{eqnarray}}

Thus, before the decay of the $X$-particles, one will have
\begin{eqnarray}
\rho_Y(\tau)=\frac{7}{4608\left( \xi-\frac{1}{6} \right)^2}\left( \frac{m_X}{m_Y} \right)^4\rho_X(\tau),
\end{eqnarray}
and, assuming that $|\xi-\frac{1}{6}|\cong 1$ so that the energy density of the $X$-particles is the maximum possible,  we will have
\begin{eqnarray}
\rho_Y(\tau)\cong 1.5\times 10^{-3}\left( \frac{m_X}{m_Y} \right)^4\rho_X(\tau).
\end{eqnarray}

Now, it is important to take into account  that, when reheating is due to the gravitational production of superheavy particles, in order that the overproduction of GWs
does not alter the  BBN success,  the decay of these particles has to take place after the end of kination \cite{hyp}. Then, assuming as usual instantaneous thermalization, the reheating is produced immediately after the decay of the $X$-particles, obtaining
\begin{eqnarray}
\rho_{Y,rh}=1.5\times 10^{-3}\left( \frac{m_X}{m_Y} \right)^4\rho_{X,rh},
\end{eqnarray}
where the subindex ``rh" means that the quantities are evaluated at the reheating time.
After reheating,  the evolution of the corresponding energy densities will be
\begin{eqnarray}
\rho_X(\tau)=\rho_{X,rh}\left(\frac{a_{rh}}{a(\tau)} \right)^4,  \rho_Y(\tau)=\rho_{Y,rh}\left(\frac{a_{rh}}{a(\tau)} \right)^3,\end{eqnarray}
meaning that at the matter-radiation equality
\begin{eqnarray}
\frac{a_{rh}}{a_{eq}}=\frac{\rho_{Y,rh}}{\rho_{X,rh}}\cong 1.5\times 10^{-3}\left( \frac{m_X}{m_Y} \right)^4,
\end{eqnarray}
and consequently
\begin{eqnarray}
\rho_{Y,eq}\cong 5\times 10^{-12}\rho_{X,rh}\left( \frac{m_X}{m_Y} \right)^{16}\nonumber\\
=
\frac{\pi^2 g_*}{6}\times 10^{-12}  T_{rh}^4 \left( \frac{m_X}{m_Y} \right)^{16},\end{eqnarray}
where $T_{rh}$ denotes the reheating temperature and $g_*=106.75$ are the degrees of freedom for the Standard Model.

On the other hand, 
 considering the central values obtained in \cite{planck} 
  of  the red-shift at the matter-radiation equality $z_{eq}=3365$,
the present value of the ratio of the matter energy density to the critical one $\Omega_{m,0}=0.308$, and $H_0=67.81\; \mbox{Km/sec/Mpc}\cong
1.42\times 10^{-33}$ eV,
one can deduce that  the present value of the matter energy density is $\rho_{m,0}=3H_0^2M_{pl}^2\Omega_{m,0}=3.26\times 10^{-121} M_{pl}^4$, and at the matter-radiation equality one will 
have $\rho_{m,eq}=\rho_{m,0}(1+z_{eq})^3=
4.4\times 10^{-1} \mbox{eV}^4$. Since practically all the matter has a non-baryonic origin, one can conclude that $\rho_{Y,eq}\cong \rho_{m,eq}$,
meaning that the reheating temperature is given by a function of $m_Y/m_X$ as follows:
{\begin{eqnarray}\label{temperature1}
{T}_{rh}\cong 2.2\times 10^{-7}\left( \frac{m_Y}{m_X} \right)^{4} \mbox{ GeV}.\end{eqnarray}}

\subsection{Decay after the end of the kination regime}
As we have already explained in the previous section, in order that the overproduction of GWs  does not alter the BBN success, the decay of the $X$-particles has to be produced after the end
of kination, which occurs when the energy density of the inflaton field is equal to the one of the $X$-particles. Then,  the decaying rate, namely $\Gamma$, has to satisfy
  ${\Gamma}\leq H(\tau_{end})\equiv H_{end}$, where we have denoted by $\tau_{end}$ the time at which kination ends.
Therefore, one has 
\begin{eqnarray}\label{31}
H^2_{end}=\frac{2\rho_{\varphi, end}}{3M_{pl}^2}, \end{eqnarray}
and \begin{eqnarray}\label{26}
\rho_{\varphi, end}=\rho_{\varphi, kin}\left( \frac{a_{kin}}{a_{end}} \right)^6=3H^2_{kin}M_{pl}^2\Theta^2,
\end{eqnarray}
in which, taking into account that during kination the energy density of the inflaton field decays as $a^{-6}$ and the one of the produced particles as $a^{-3}$,  we have introduced the so-called {\it heating efficiency} defined in \cite{rubio} as
{
\begin{eqnarray}\label{27}\Theta\equiv \left( \frac{a_{kin}}{a_{end}} \right)^3=
\frac{\rho_{X,kin}}{\rho_{\varphi, kin}}\cong 5\times 10^{-34}\left(\frac{M_{pl}}{m_X} \right)^4.\end{eqnarray}

Consequently,  (\ref{31}) leads to  $H_{end}=\sqrt{2}H_{kin}\Theta$, and 
from the constraint $\Gamma\leq H_{end}$
one obtains the bound
\begin{eqnarray}\label{const1}
\frac{\Gamma}{M_{pl}}\leq  10^{-39}\left(\frac{M_{pl}}{m_X} \right)^4.\end{eqnarray}

}


On the other hand,  assuming  once again instantaneous thermalization,  
the energy density of the $X$-particles at the reheating time will be $\rho_{ X,rh}=3{\Gamma}^2M_{pl}^2$, and thus,
the reheating temperature  will be
given by
\begin{eqnarray}\label{temperature2}
T_{rh}=
\left( \frac{90}{\pi^2 g_*} \right)^{\frac{1}{4}}\sqrt{{\Gamma}M_{pl}}
\cong 1.3 \times 10^{18} \sqrt{\frac{\Gamma}{M_{pl}}} \mbox{ GeV}.
\end{eqnarray}

As a consequence, from the two expressions of the reheating temperature (\ref{temperature1}) and (\ref{temperature2}) one can write the mass of the dark matter
as a function of $\Gamma$ and $m_X$ as follows:
{\begin{eqnarray}\label{darkmatter}
m_Y\cong 1.55\times 10^6\left(\frac{\Gamma}{M_{pl}} \right)^{1/8}m_X.
\end{eqnarray}}

\subsection{Overproduction of GWs }
\label{sec-overproduction}
The success of the BBN demands that the ratio of the energy density of GWs to the one of the produced particles at the reheating time satisfies \cite{hossain3}
\begin{eqnarray}\label{bbnconstraint}
\frac{\rho_{GW, rh}}{\rho_{X,rh}}\leq 10^{-2},
\end{eqnarray} 
where the energy density of the GWs is given by $\rho_{GW}(\tau)\cong 10^{-2} H^4_{kin} \left(\frac{a_{kin}}{a(\tau)} \right)^4$  (see for instance \cite{ford}).

Thus, since 
\begin{eqnarray}
\rho_{GW, rh}=
10^{-2} H^4_{kin}
\left( \frac{\Gamma}{\sqrt{2\Theta}H_{kin} }  \right)^{8/3},
\end{eqnarray}
we will have, \begin{eqnarray}
\frac{\rho_{GW, rh}}{\rho_{X,rh}}\cong 5.4\times 10^{33}\left( \frac{m_X}{M_{pl}} \right)^{16/3}\left( \frac{\Gamma}{M_{pl}} \right)^{2/3},
\end{eqnarray}
meaning that
the bound (\ref{bbnconstraint}) leads to the constraint
\begin{eqnarray}\label{const2}
\frac{\Gamma}{M_{pl}}\leq 2.5 \times 10^{-54}\left(\frac{M_{pl}}{m_X} \right)^8.
\end{eqnarray}

{
Here, it is important to recall that, in order to apply the WKB approximation, we have assumed that the mass of the particles is greater than the Hubble parameter
at the beginning of inflation, which is of the order of $10^{14}$ GeV if inflation starts at GUT scales. Therefore, choosing $m_X\geq 10^{15}$ GeV one can easily show
that
the constraint 
(\ref{const2}) automatically implies (\ref{const1}), and thus, taking into account that $T_{rh}> 1$ MeV because the BBN occurs at the MeV regime \cite{gkr}, one gets that
$\Gamma$ must satisfy 
\begin{eqnarray}\label{const3} 
5.9\times 10^{-43}\leq
\frac{\Gamma}{M_{pl}}\leq 2.5 \times 10^{-54}\left(\frac{M_{pl}}{m_X} \right)^8,
\end{eqnarray}
which always holds when 
\begin{eqnarray}
 10^{15} \mbox{ GeV}\leq m_X\leq 9 \times 10^{16} \mbox{ GeV}.
\end{eqnarray}

Consequently, from (\ref{temperature2}) and (\ref{const3}), for our model  the reheating temperature is bounded by
\begin{eqnarray}
1 \mbox{ MeV}\leq T_{rh}\leq 
2\times 10^{-9} \left( \frac{M_{pl}}{m_X}\right)^4 \mbox{ GeV},
\end{eqnarray}
and  from   (\ref{darkmatter}) and (\ref{const3}) the mass of the $Y$-particles by
{\begin{eqnarray}
8.16 m_X
\leq m_Y\leq 7.4 \times 10^{17}\mbox{ GeV}.
\end{eqnarray}}

}

Then, choosing for example $m_X=10^{15}$ GeV, one gets the following bound for the reheating temperature
\begin{eqnarray}
1 \mbox{ MeV}\leq T_{rh}\leq 
66 \mbox{ TeV},
\end{eqnarray}
and from (\ref{temperature1}), if one assumes that the universe reheats when the temperature is around $1$ GeV, the mass of
the $Y$-particles has to be $m_Y\cong 4.6\times 10^{16} $ GeV.  In general, for $m_X=10^{15}$ GeV the relation between the mass of the particles that
generate dark matter and the reheating temperature is presented in Figure \ref{fig:temperature}.

\


\begin{figure}[H]
\begin{center}
\includegraphics[scale=0.45]{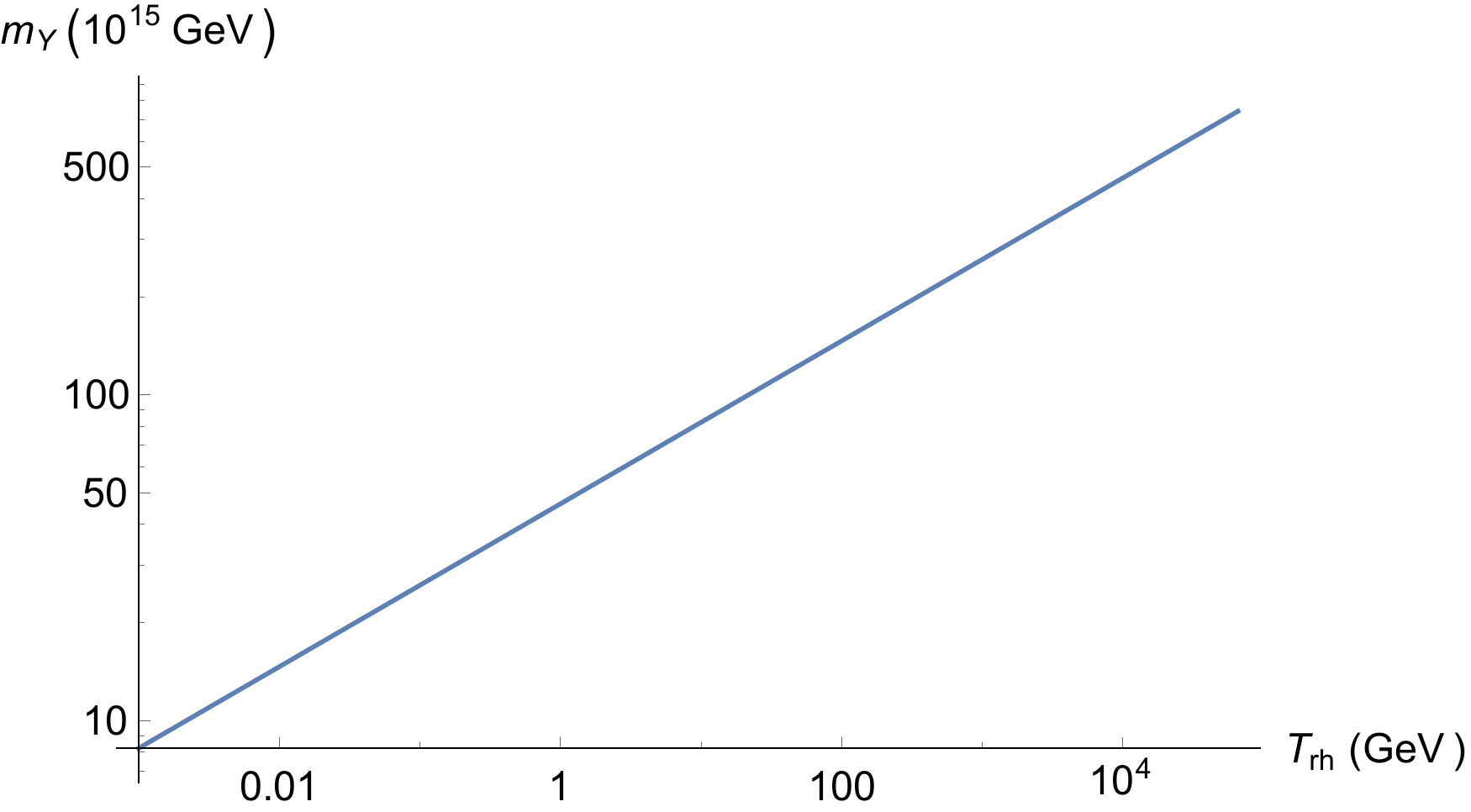}
\end{center}
\caption{Mass of the $Y$-particles as a function of the reheating temperature.}
\label{fig:temperature}
\end{figure}

\

To end this section, a final remark is in order: When one considers that the $X$-particles are conformally coupled with gravity, the relation (\ref{temperature1})
becomes
\begin{eqnarray}
T_{rh}\cong 3.3 \times 10^{-10}\left( \frac{m_Y}{m_X} \right)^4 \mbox{ GeV}.
\end{eqnarray}
Then, for $m_X=10^{15}$ GeV and a reheating temperature of $1$ GeV, one gets $m_Y\cong 2.3\times 10^{17} $ GeV, which means that the mass of the $Y$- particles is
increased in one order with respect to the nonconformally coupled case. This shows that, if one wants a model with elementary  superheavy $X$ and $Y$ far from the Planck
scale, one has to consider that the $X$-particles -the ones which decay into light baryonic matter- do not have to be conformally coupled with gravity.

\

\section{Numerical calculations}
In this section we want to calculate the value of the parameter $M$ as a function of the reheating temperature and the late time evolution of our model.

\subsection{Analytic results}
To perform this calculation, first of all, as we have already shown at the end of  Section $2$,  we take as initial conditions at the beginning of kination 
\begin{eqnarray}
\varphi_{kin}=0, \quad \dot{\varphi}_{kin}=3.54\times 10^{-6} M_{pl}^2.
\end{eqnarray}

During kination one can safely disregard the potential, so
 during this phase one has $a\propto t^{1/3}\Longrightarrow H=\frac{1}{3t}$, and using the Friedmann equation, the dynamics in this regime will be
\begin{eqnarray}
\frac{\dot{\varphi}^2}{2}=\frac{M_{pl}^2}{3t^2}\Longrightarrow \dot{\varphi}=\sqrt{\frac{2}{3}}\frac{M_{pl}}{t}\nonumber \\ \Longrightarrow 
\varphi(t)=\sqrt{\frac{2}{3}}M_{pl}\ln \left( \frac{t}{t_{kin}} \right).\end{eqnarray}

Then, 
at the end of kination, one has 
\begin{eqnarray}
\varphi_{end}=\sqrt{\frac{2}{3}}M_{pl}\ln\left( \frac{H_{kin}}{H_{end}} \right), 
\dot{\varphi}_{end}=\sqrt{6}M_{pl}H_{end},
\end{eqnarray}
and using once again that $H_{end}=\sqrt{2}H_{kin}\Theta$, one gets
 \begin{eqnarray}
\varphi_{end}=-\sqrt{\frac{2}{3}}M_{pl}\ln\left( \sqrt{2}\Theta \right), 
\dot{\varphi}_{end}=2\sqrt{3}M_{pl}H_{kin}\Theta.
\end{eqnarray}

During the period between $t_{end}$ and $t_{rh}$ the universe is matter dominated and, thus, the Hubble parameter becomes $H=\frac{2}{3t}$. Since the gradient of the potential could also be disregarded at this
epoch,  hence, the equation of the scalar field becomes $\ddot{\varphi}+\frac{2}{t}\dot{\varphi}=0$, and thus, at the reheating time
\begin{eqnarray}
\varphi_{rh}=\varphi_{end}+\sqrt{\frac{2}{3}}M_{pl}\left( 1-\frac{{t}_{end}}{t_{rh}}  \right)\nonumber\\
=\varphi_{end}+\sqrt{\frac{2}{3}}M_{pl}\left( 1-\frac{H_{rh}}{2H_{end}}\right)\nonumber\\ =
\varphi_{end}+\sqrt{\frac{2}{3}}M_{pl}\left( 1-
{\frac{\pi}{6}\sqrt{\frac{g_{*}}{10}}\frac{T_{rh}^2}{M_{pl} H_{kin}\Theta}}\right),\end{eqnarray}
and
\begin{eqnarray}
\dot{\varphi}_{rh}
=\frac{\sqrt{3}}{4}\frac{M_{pl}H_{rh}^2}{H_{kin}\Theta}
=\frac{\sqrt{3}\pi^2}{180}\frac{g_{*}T_{rh}^4}{H_{kin}M_{pl}\Theta}.\end{eqnarray}

Note that for the allowed reheating temperatures, i.e., for temperatures satisfying $T_{rh}\leq 66$ TeV, one has 
${\frac{\pi}{6}\sqrt{\frac{g_{*}}{10}}\frac{T_{rh}^2}{M_{pl} H_{kin}\Theta}}\ll 1$, so we can safely make the approximation
\begin{eqnarray}
\varphi_{rh}\cong \varphi_{end}+\sqrt{\frac{2}{3}}M_{pl}.
\end{eqnarray}

During the radiation period one can continue disregarding the potential,  obtaining
\begin{eqnarray}
\varphi(t)=\varphi_{rh}+2\dot{\varphi}_{rh}t_{tr}\left(1-\sqrt{\frac{t_{rh}}{t}}\right),
\end{eqnarray}
and thus, at the matter-radiation equality one has
\begin{align}\label{eq}
 \varphi_{eq} =\varphi_{rh}+2\sqrt{\frac{2}{3}}M_{pl}\left(1-\sqrt{\frac{4H_{eq}}{3H_{rh}}}\right)\nonumber\\
 =\varphi_{rh}+2\sqrt{\frac{2}{3}}M_{pl}\left(1-\sqrt{\frac{4}{3}}\left( \frac{g_{eq}}{g_{*}} \right)^{1/4}\frac{T_{eq}}{T_{rh}}\right) \nonumber \\
\cong \varphi_{rh}+2\sqrt{\frac{2}{3}}M_{pl},
  \end{align}
 where $g_{eq}\cong 3.36$ are the degrees of freedom at this scale  \cite{gr} and $T_{eq}$ is the temperature of the radiation at the matter-radiation equilibrium, which is related with the energy
 density via the relation $\rho_{eq}=\frac{\pi^2}{15}g_{eq}T^4_{eq}$, and thus, given by $T_{eq}\cong 7.8\times 10^{-10}$ GeV.
 
In the same way,  
 \begin{align}\label{doteq}
\dot{\varphi}_{eq}=\dot{\varphi}_{rh}\frac{t_{rh}}{t_{eq}}\sqrt{\frac{t_{rh}}{t_{eq}}}=\left(\frac{16g_{eq}}{9g_*}\right)^{3/4}\left(\frac{T_{eq}}{T_{rh}}\right)^3 \dot{\varphi}_{rh}.
\end{align}

\

After the matter-radiation equality the dynamical equations can not be solved analytically and, thus, one needs to use numerics to compute them. In order to do that, we need to use a ``time'' variable  that we choose to be the number of $e$-folds up to the present epoch, namely, $N\equiv -\ln(1+z)=\ln\left( \frac{a}{a_0}\right)$. Now, using  the variable $N$,  one can recast the  energy density of radiation and matter respectively as
\begin{eqnarray}
\rho_{r}(a)=\frac{\rho_{eq}}{2}\left(\frac{a_{eq}}{a}  \right)^4\Longrightarrow \rho_{r}(N)= \frac{\rho_{eq}}{2}e^{4(N_{eq}-N)} ,
\end{eqnarray}
and
\begin{eqnarray}
\rho_{m}(a)=\frac{\rho_{eq}}{2}\left(\frac{a_{eq}}{a}  \right)^3\Longrightarrow \rho_{m}(N)=\frac{\rho_{eq}}{2}e^{3(N_{eq}-N)},
\end{eqnarray}
where  the value of the energy density at the matter-radiation equality { $\rho_{eq}\cong 8.8\times 10^{-1}$  ${\mbox eV}^4$}
has been obtained in the previous Section III and one can also understand that $N_{eq}$ is the value of $N$ at the matter-radiation equality.

\begin{figure*}[ht]
\begin{center}
\includegraphics[scale=0.5]{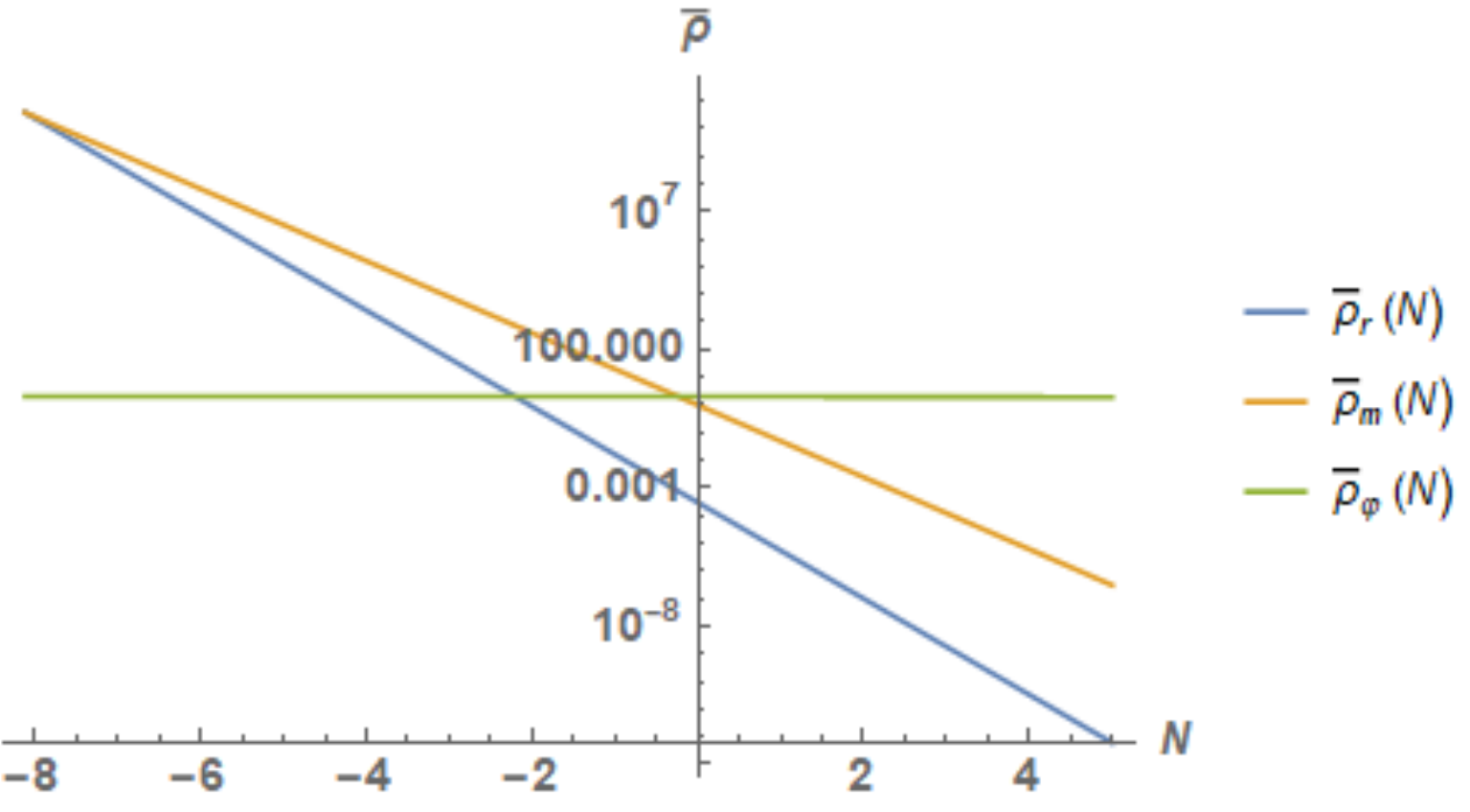}
\includegraphics[scale=0.5]{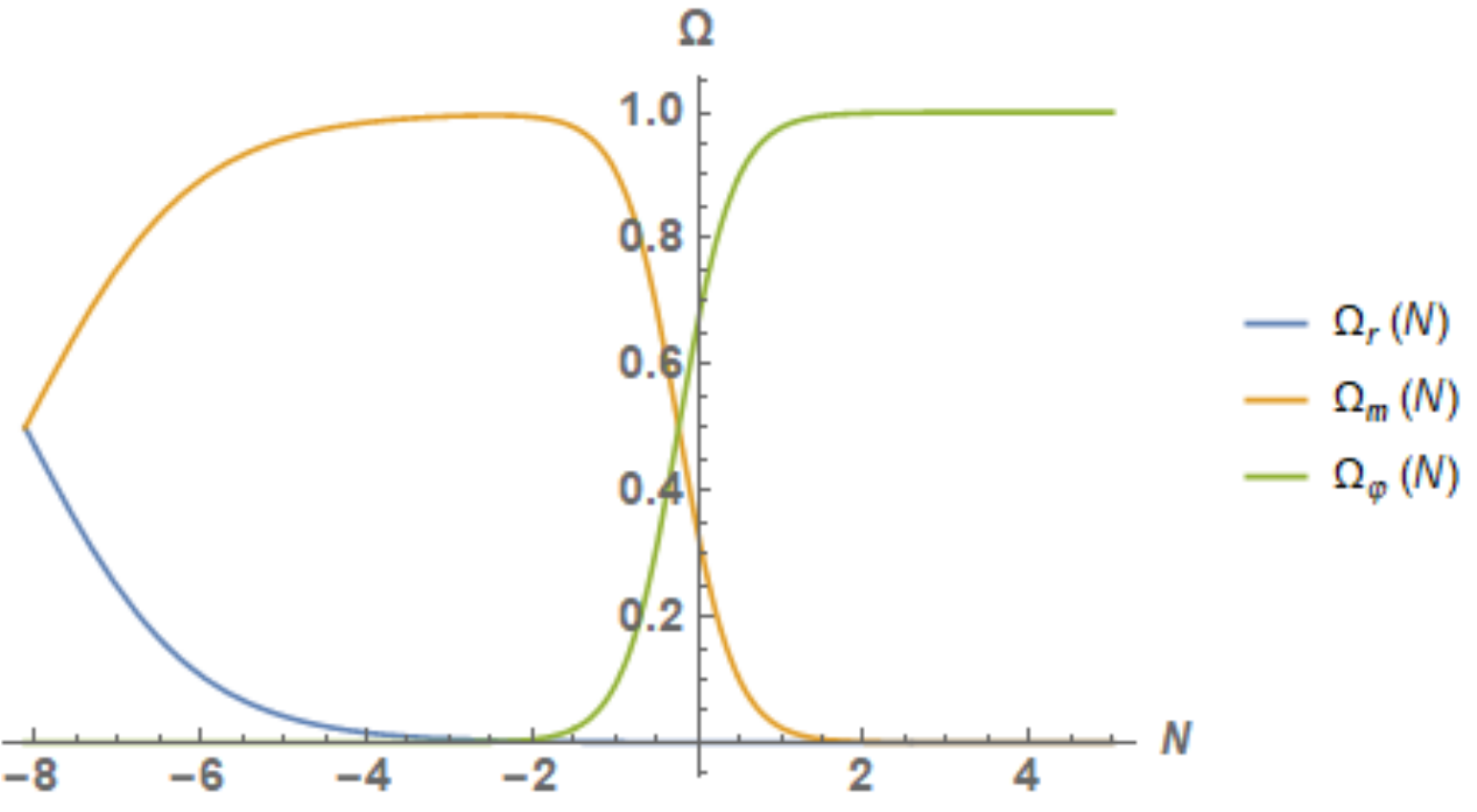}
\end{center}
\caption{Evolution of $\{\bar{\rho}_B(N)\}_{B=r,m,\varphi}$ and $\{\Omega_B(N)\}_{B=r,m,\varphi}$ .}
\label{fig:densitats}
\end{figure*}

\begin{figure}[ht]
\begin{center}
\includegraphics[scale=0.45]{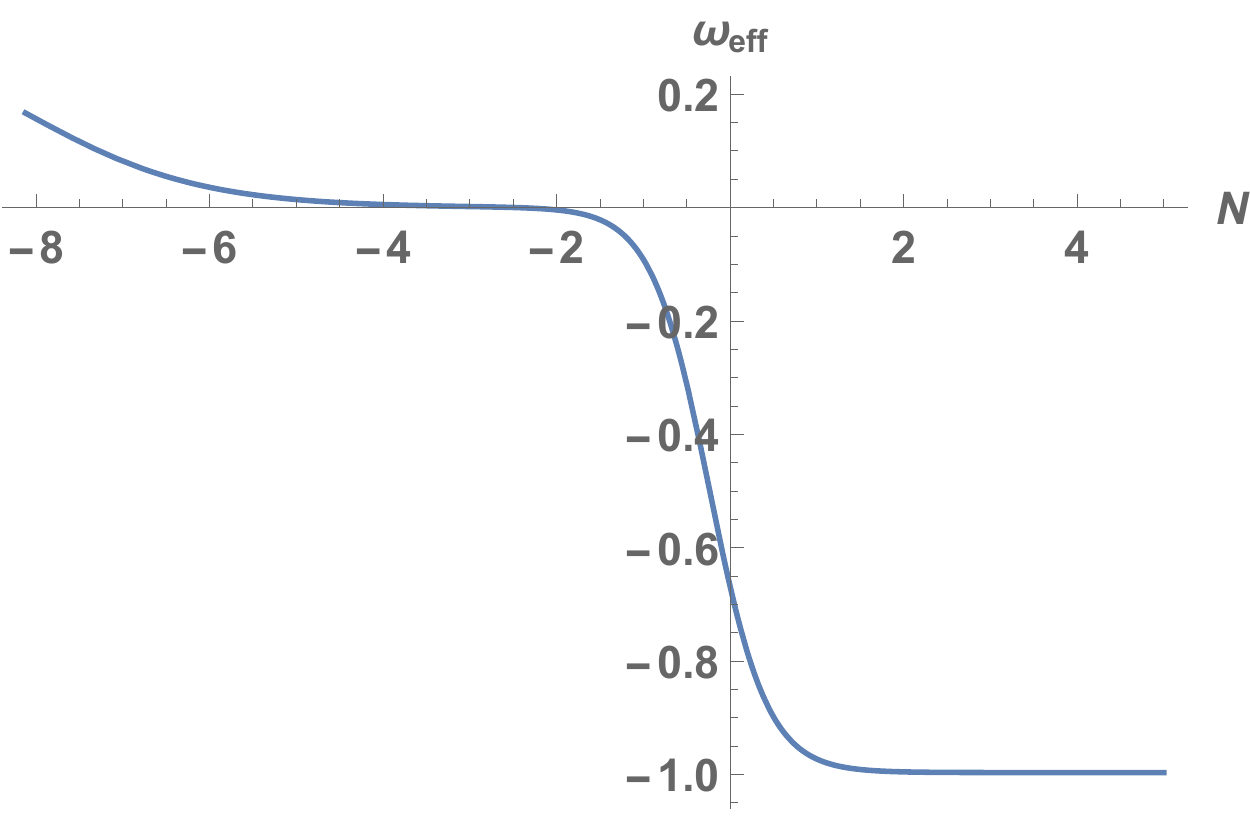}
\end{center}
\caption{Evolution of the effective EoS parameter.}
\label{fig:weff}
\end{figure}

\subsection{The dynamical system}
In order to obtain the dynamical system for this scalar field model, we 
introduce the following dimensionless variables
 \begin{eqnarray}
 x=\frac{\varphi}{M_{pl}}, \qquad y=\frac{\dot{\varphi}}{H_0 M_{pl}},
 \end{eqnarray}
 where $H_0\sim 1.42 \times 10^{-33}$ eV denotes once again the current value of the Hubble parameter. Now, using the variable
 $N = - \ln (1+z)$ defined above and also using the conservation equation $\ddot{\varphi}+3H\dot{\varphi}+V_{\varphi}=0$, one can construct the following  non-autonomous dynamical system:
 \begin{eqnarray}\label{system}
 \left\{ \begin{array}{ccc}
 x^\prime & =& \frac{y}{\bar H}~,\\
 y^\prime &=& -3y-\frac{\bar{V}_x}{ \bar{H}}~,\end{array}\right.
 \end{eqnarray}
 where the prime represents the derivative with respect to $N$, $\bar{H}=\frac{H}{H_0}$   and $\bar{V}=\frac{V}{H_0^2M_{pl}^2}$. Moreover, the Friedmann equation now looks as  
 \begin{eqnarray}\label{friedmann}
 \bar{H}(N)=\frac{1}{\sqrt{3}}\sqrt{ \frac{y^2}{2}+\bar{V}(x)+ \bar{\rho}_{r}(N)+\bar{\rho}_{m}(N) }~,
 \end{eqnarray}
where we have introduced the following dimensionless energy densities
 $\bar{\rho}_{r}=\frac{\rho_{r}}{H_0^2M_{pl}^2}$ and 
 $\bar{\rho}_{m}=\frac{\rho_{m}}{H_0^2M_{pl}^2}$.

Then, we have to integrate the dynamical system, starting at $N_{eq}=-8.121$, with initial condition $x_{eq}$ and $y_{eq}$ which are obtained analytically in the previous subsection. The value of the parameter $M$ is obtained equaling at $N=0$ the equation (\ref{friedmann}) to $1$, i.e., imposing $\bar{H}(0)=1$.

\

Numerical calculations show that $M\cong 2.6\times 10^5$ GeV, independently of the reheating temperature, which is a value of the same order as the one obtained in \cite{pv}.
On the other hand, in Figure \ref{fig:densitats} we have drawn the evolution of the different dimensionless energy densities, obtaining a frozen quintessence, that is,
the energy density of the scalar field is frozen and starts to dominate very close to the present time.
We have also considered the evolution of the ratio of the energy density to the critical one for the different constituents, i.e., $\Omega_{B}(t)=\frac{\rho_{B}(t)}{3H^2(t)M_{pl}^2}$
where $B=r,m,\varphi$. And in Figure \ref{fig:weff} we have depicted 
the evolution of the effective EoS parameter $\omega_{eff}(t)=-1-\frac{2\dot{H}(t)}{3H^2(t)}$. Finally, we can see that for $N$  greater than $1.5$ one has  $\Omega_{\varphi}=1$ and $\omega_{eff}=-1$
meaning that, at late times, the universe enters in a de Sitter phase and, thus, exhibits an eternal acceleration.

\section{Conclusions}
In the present work we have considered a quintessential model whose potential, which only depends on two parameters, is composed by a Starobinsky Inflationary type-potential matched with an inverse power law potential, which is responsible for quintessence. Since the phase transition from the end of inflation to the beginning of kination is very abrupt, the adiabatic regime is broken and particles are produced. We have assumed that during this period two kind of superheavy particles are gravitationally produced: $X$-particles, which are nonconformally coupled with gravity and whose decay products form the baryonic matter, and $Y$-particles, which are conformally coupled with gravity but are only GIMP, and thus, they are responsible for the dark matter abundance. For this model we have shown that, for reasonable masses of the $X$-particles around $10^{15}$ GeV, a viable model with a reheating temperature in the GeV regime is obtained when the mass of the dark matter particles is of the order of $5\times 10^{16}$ GeV. Finally, we have shown numerically that the model leads, at late times,  to a frozen quintessence, and thus, to an eternal inflation.

\

{\it Acknowledgments.}   
 We want to thank Prof. Salvatore Capozziello for telling us, during the workshop Modified Gravity and Cosmology,  the possibility to consider the production of superheavy particles 
nonconformally coupled with gravity in order to reduce the masses of the particles involved in the theory. This investigation has been supported by MINECO (Spain) grant
 MTM2017-84214-C2-1-P, and  in part by the Catalan Government 2017-SGR-247.

\

\section*{Appendix:  The number of e-folds in quintessential inflation}

In this Appendix we will perform an accurate calculation of the number of e-folds for our model, i.e., for a quintessential inflation model where reheating is produced after the end of kination, and we will see that, due to the kination era, the number of e-folds is greater than in standard inflation, where reheating is produced due the oscillations of the
inflaton field (see for a detailed calculation \cite{Cook}).

\

Let $k_*=a_*H_*$ be the value of the pivot scale in co-moving coordinates when it exits the Hubble radius and $N$ the number of e-folds from the exiting of the pivot scale to the end of inflation, i.e., $a_{END}=a_*e^N$, where, once again, $a_{END}$ denotes the value at the end of inflation.

Now we write
\begin{eqnarray}
\frac{k_*}{a_0H_0}=e^{-N}\frac{H_*}{H_0}\frac{a_{END}}{a_{kin}}\frac{a_{kin}}{a_{end}}\frac{a_{end}}{a_{rh}}\frac{a_{rh}}{a_{eq}}\frac{a_{eq}}{a_{0}},
\end{eqnarray}
where, as in previous sections, $a_{kin}$, $a_{end}$, $a_{rh}$, $a_{eq}$ and $a_0$ denote the value of the scale factor at the beginning of kination, at the end of
kination, at the reheating time, at the matter-radiation equality and at present time, respectively.

Choosing, as usual, $k_{phys}\equiv \frac{k_*}{a_0}=0.02 \mbox{Mpc}^{-1}$ and taking into account that $H_0\cong 2\times 10^{-4} \mbox{Mpc}^{-1}$ one gets
\begin{eqnarray}
10^2=e^{-N}\frac{H_*}{H_0}\frac{a_{END}}{a_{kin}}
\frac{\rho_{X,rh}^{1/3}}{\rho_{\varphi,end}^{1/6} \rho_{\varphi,kin}^{1/6} }\left( \frac{\rho_{X,eq}}{\rho_{X,rh}}\right)^{1/4}
\frac{a_{eq}}{a_{0}},
\end{eqnarray}
where we have used that 
\begin{eqnarray}
\rho_{\varphi,end}=\rho_{\varphi,kin}\left(\frac{a_{kin}}{a_{end}}  \right)^6, \rho_{X,rh}=\rho_{\varphi,end}\left(\frac{a_{end}}{a_{rh}}  \right)^3\nonumber \\
\mbox{ and } \rho_{X,eq}=\rho_{X,rh}\left(\frac{a_{rh}}{a_{eq}}  \right)^4.
\end{eqnarray}

Now, using that $\frac{\rho_{X,eq}}{\rho_{X,rh}}
=\frac{g_{eq}T^4_{eq}}{g_* T^4_{rh}}$, 
where, as we have already seen in the previous Section, the number of degrees of freedom at the matter-radiation equality is $g_{eq}=3.36$, and taking into account that after reheating the evolution is adiabatic, i.e., $a_{eq}T_{eq}=a_0T_0$, one gets
\begin{eqnarray}
10^2=e^{-N}\frac{H_*}{H_0}\frac{a_{END}}{a_{kin}}
\frac{\rho_{X,rh}^{1/3}}{\rho_{\varphi,end}^{1/6} \rho_{\varphi,kin}^{1/6} }\left( \frac{g_{eq}}{g_*}\right)^{1/4}
\frac{T_0}{T_{rh}},
\end{eqnarray}
and, from equations (\ref{8}) and (\ref{26}) and using that $\rho_{X,rh}=\frac{\pi^2}{30}g_*T_{rh}^4$, we obtain
\begin{eqnarray}
10^{-2}\cong 0.75e^{-N}\frac{H_*}{H_0}\frac{a_{END}}{a_{kin}}
\left( \frac{T_{rh}}{M_{pl}}\right)^{4/3}\Theta^{-1/3}
\frac{T_0}{T_{rh}}.
\end{eqnarray}

At this point, we use the observational data $T_0\cong 2.33\times 10^{-13}$ GeV and $H_0\cong 5.95\times 10^{-61} M_{pl}$ to get
\begin{eqnarray}
e^N\cong 9.1\times 10^{24} \frac{H_*}{M_{pl}}\frac{a_{END}}{a_{kin}}
\left( \frac{T_{rh}}{\mbox{GeV}} \right)^{1/3}\Theta^{-1/3},
\end{eqnarray}
and finally, from equation (\ref{27}) and the value $H_*\cong \sqrt{0.3}\pi (1-n_s)M_{pl}\times 10^{-4}$, we conclude that
\begin{eqnarray}
e^N\cong 2\times 10^{32} (1-n_s)\frac{a_{END}}{a_{kin}}
\left( \frac{T_{rh}}{\mbox{GeV}} \right)^{1/3}\left( \frac{m_X}{M_{pl}}\right)^{4/3}.
\end{eqnarray}

Thus, since for many solvable models \cite{haro1,haro2} one has $\ln\left(\frac{a_{END}}{a_{kin}} \right)\cong -1$, we obtain
\begin{eqnarray}
N\cong 73.37+\ln(1-n_s)+\frac{1}{3}\ln\left( \frac{T_{rh}}{\mbox{GeV}} \right)+\frac{4}{3}\ln\left( \frac{m_X}{M_{pl}}\right).\end{eqnarray}

In particular, for $m_X=10^{15}$ GeV and $n_s=0.968$, one gets
\begin{eqnarray}
N\cong 59.55+\frac{1}{3}\ln\left( \frac{T_{rh}}{\mbox{GeV}} \right),\end{eqnarray}
which for the allowed temperatures $1 \mbox{ MeV}\leq T_{rh}\leq 66 \mbox{ TeV}$ (see formula (44)), leads to 
\begin{eqnarray}
57.25\leq N\leq 63.25.
\end{eqnarray}

\end{document}